\documentclass[leqno,twocolumn]{article}
\usepackage{ltexpprt}

\pdfoutput=1

\usepackage{ifpdf}
\ifpdf
\usepackage{graphicx}
\usepackage[bookmarks=false]{hyperref}
\else
\usepackage[dvips]{graphicx}
\usepackage[dvips,bookmarks=false]{hyperref}
\fi

\usepackage[algo2e,vlined]{algorithm2e}
\usepackage{amssymb, amsmath}

\newcommand{\clust}{\ensuremath{\mathcal{C}}}
\newcommand{\bigO}{\ensuremath{\mathcal{O}}}

\begin{document}

\title{Multi-Level Algorithms for Modularity Clustering}
\author{
  Andreas Noack\footnotemark \addtocounter{footnote}{-1}
    \and
  Randolf Rotta\thanks{
    Brandenburg University of Technology, 03013 Cottbus, Germany;
    {\tt \{an,rrotta\}@informatik.tu-cottbus.de}
  }
}
\date{}

\maketitle

\begin{abstract} \small\baselineskip=9pt
Modularity is one of the most widely used quality measures
for graph clusterings.
Maximizing modularity is NP-hard, and the runtime of exact algorithms
is prohibitive for large graphs.
A simple and effective class of heuristics coarsens
the graph by iteratively merging clusters (starting from
singletons), and optionally refines the resulting clustering
by iteratively moving individual vertices between clusters.
Several heuristics of this type have been proposed in the literature,
but little is known about their relative performance.

This paper experimentally compares existing and new
coarsening- and refinement-based heuristics with respect to
their effectiveness (achieved modularity) and efficiency (runtime).
Concerning coarsening, it turns out that the most widely
used criterion for merging clusters (modularity increase)
is outperformed by other simple criteria,
and that a recent algorithm by Schuetz and Caflisch
is no improvement over simple greedy coarsening for these criteria.
Concerning refinement, a new multi-level algorithm is shown to produce
significantly better clusterings than conventional single-level algorithms.
A comparison with published benchmark results and algorithm implementations
shows that combinations of coarsening and multi-level refinement
are competitive with the best algorithms in the literature.

\end{abstract}

\section{Introduction}
A {\em graph clustering} partitions the vertex set of a graph
into disjoint subsets called {\em clusters}.
{\em Modularity} was introduced by Newman and Girvan as formalization
of the common requirement that the connections within graph clusters
should be dense, and the connections between different graph clusters
should be sparse~\cite{newman2004fae}.
It is by far not the only quality measure for graph clusterings
\cite{Gaertler:2005,Schaeffer:CSR2007},
but one of the most widely used measures,
and has been successfully applied for detecting meaningful groups
in a wide variety of complex systems.

The problem of finding a clustering with maximum modularity for a given graph
is NP-hard~\cite{BrandesEtAl:TKDE2008}, and even recent exact algorithms
scale only to graphs with a few hundred vertices
\cite{BrandesEtAl:TKDE2008,agarwal2008mmg,XuEtAl:EPJ2007}.
In practice, modularity is almost exclusively optimized
with heuristic algorithms.
Like modularity itself, many of these heuristics have been proposed
in the physics literature.

A particularly simple heuristic is the iterative merging of cluster pairs,
starting from singleton clusters, and
always choosing the merge that results in the largest modularity increase.
This {\em greedy coarsening} can be efficiently implemented
and produces reasonable clusterings \cite{newman2004fad,clauset2004vln},
but was soon observed to be biased towards merging large clusters
\cite{danon2006esh,wakita2007msn}.
To remove this bias and obtain clusterings with even higher modularity,
researchers suggested to replace modularity increase with other
prioritizing criteria for potential merges \cite{danon2006esh,wakita2007msn},
and to modify the purely greedy merge strategy \cite{schuetz2008emo}.
However, the numerous proposals have not been organized
into a coherent design space, and the published evaluation results are largely
incomparable due to the use of different (and often small) graph collections.
Therefore, Section~\ref{s:coarsening} systematically describes
major design alternatives for coarsening algorithms,
including two new prioritizing criteria for merges,
and Section~\ref{ss:ecoarsening} compares them experimentally.

The clusterings produced by coarsening heuristics can be improved
with {\em refinement algorithms}, which iteratively move individual vertices
between clusters~\cite{schuetz2008emo}.
An obvious solution is greedy refinement, which always chooses the vertex move
resulting in the largest increase of modularity.
However, moving the vertices in arbitrary order (instead of always moving
the best vertex) is much faster and not necessarily less effective,
and adaptations of the classic Kernighan-Lin refinement~\cite{kernighan1970ehp}
are not much slower and have some capability to escape local maxima.
All of these algorithms can be applied not only to the original graph,
but to any level of the coarsening hierarchy,
by considering each cluster of the coarsening level as a single coarse vertex.
This {\em multi-level refinement} is extremely effective
for minimum cut partitioning problems~\cite{hendrickson1995mlp,karypis1998fah},
but has not previously been adapted to modularity clustering.
Section~\ref{s:refinement} details the single-level and multi-level refinement
heuristics, and Section~\ref{ss:erefinement} compares them experimentally.
Because the effectiveness of (particularly multi-level) refinement
may depend on the coarsening algorithm, Section~\ref{ss:ecombi} examines
various combinations of coarsening and refinement heuristics.

Section~\ref{s:related} compares public implementations
and benchmark results of modularity clustering heuristics,
without a restriction to coarsening and refinement algorithms.
While this is one of the most extensive comparisons in the literature,
it is far from exhaustive, because implementations and sufficient
experimental results have not been published for some proposed heuristics.
The main purpose is to demonstrate that particular combinations
of simple coarsening and multi-level refinement algorithms
are (at least) competitive with the best available heuristics, and thus
the results of the previous sections are indeed practically significant.

\section{\label{s:problem}Graph Clusterings and Modularity}

\subsection{\label{ss:graph}Graph Clusterings.}
A {\em graph} $(V,f)$ consist of a finite set~V of {\em vertices}
and a function~$f: V \times V \to \mathbb{N}$ that assigns a nonnegative
{\em edge weight} to each vertex pair.
For simplicity, graphs are assumed to be undirected, i.e., $f(u,v) = f(v,u)$
for all $u,v \in V$.
The {\em degree}~$\deg(v)$ of a vertex~$v$ is the total weight
$\sum_{u \in V} f(u,v)$ of its edges.
The degrees and weights are naturally generalized
to sets of vertices, e.g., $f(V,V) = \sum_{u \in V, v \in V} f(u,v)$.
Note that $\deg(V) = f(V,V)$.

A {\em graph clustering} $\clust = \{ C_1, \dots, C_k \}$
partitions the vertex set~$V$ into disjoint non-empty subsets~$C_i$.

\subsection{\label{ss:modularity}Modularity.}
Modularity is a quality measure for graph clusterings.
It was originally introduced for graphs
where the edge weights are either $0$ or~$1$~\cite{newman2004fae},
and was later generalized to arbitrary edge weights~\cite{Newman:2004}.
The {\em modularity} of a clustering~$\clust$ is defined as
\begin{equation*}
Q(\clust) ~:=~ \sum_{C \in \clust} \left(
    \frac{   f(C,C)}{f(V,V)}
  - \frac{\deg(C)^2}{\deg(V)^2}
\right).
\end{equation*}

Intuitively, the first term is the {\em actual} fraction
of intra-cluster edge weight.
In itself, it is not a good measure of clustering quality, because it takes
the maximum value~$1$ for the trivial clustering where one cluster contains
all vertices.
The second term specifies the {\em expected} fraction
of intra-cluster edge weight in a null model
where the end-vertices of $\frac{1}{2}\deg(V)$ edges are chosen at random,
and the probability that an end-vertex of an edge attaches to
a particular vertex~$v$ is $\frac{\deg(v)}{\deg(V)}$~\cite{newman2006fcs}.
In this null model, the edge weight $f(u,v)$
between each vertex pair $(u,v) \in V^2$ is binomially distributed
with the expected value $\frac{\deg(u)\deg(v)}{\deg(V)}$.

It can be easily verified that merging two clusters $C$ and~$D$
increases the modularity by
\[
\Delta Q_{C,D} ~:=~ \frac{2f(C,D)        }{f(V,V)}
               ~-~  \frac{2\deg(C)\deg(D)}{\deg(V)^2}~,
\]
and moving a vertex~$v$ from its current cluster~$C$ to another cluster~$D$
increases the modularity by
\begin{eqnarray*}
\Delta Q_{v \to D} & := & \frac{2f(v,D)         - 2f(v,C\!-\!v)        }{f(V,V)} \\
                   & -  & \frac{2\deg(v)\deg(D) - 2\deg(v)\deg(C\!-\!v)}{\deg(V)^2}~.
\end{eqnarray*}

\section{\label{s:coarsening}Coarsening Algorithms}

Greedy coarsening algorithms iteratively merge either one cluster pair,
as detailed in the first subsection, or several disjoint cluster pairs,
as detailed in the second subsection, and choose the merged cluster pairs
according to certain priority criteria, which are discussed
in the third subsection.

\subsection{Single-Step Greedy.}\label{ss:singlestep}

The Single-Step Greedy algorithm starts with single-vertex clusters,
and iteratively merges the cluster pair with the highest priority,
until this merge would decrease the modularity.

\begin{algorithm2e}
  \SetKwData{graph}{graph}
  \SetKwData{clustering}{clustering}
  \SetKwData{ranker}{merge prioritizer}
  \SetKwData{merged}{merged}
  \SetKwData{MP}{MP}
  \SetKwData{source}{source}
  \SetKwData{target}{target}
  \SetKw{brk}{break}
  \SetKw{True}{true}
  \SetKw{False}{false}

  \Titleofalgo{~Single-Step Greedy Coarsening}
  \KwIn{\graph, \ranker}
  \KwOut{\clustering}
  \BlankLine

  initialize \clustering with singleton clusters\;
  \While{{\rm prioritized pair} $(C,D)$ {\rm satisfies} $\Delta Q_{C,D}>0$}{
    merge clusters $C$ and $D$\;
  }
\end{algorithm2e}

\paragraph{Implementation and Runtime.}

For the calculation of the priorities (see Section~\ref{ss:priority})
it is necessary to quickly retrieve the total edge weights
between adjacent clusters.  These total weights change
locally with each merge and are thus stored in a dynamically coarsened
graph where each cluster is represented by a single vertex.
In each merge of two vertices $u$ and~$v$, the edge list of the vertex
with fewer edges (say~$u$) is merged into the edge list of the other vertex.
Using the sorted double-linked edge lists proposed by
Wakita and Tsurumi~\cite{wakita2007msn},
this requires linear time in the list lengths.
However, if some neighbor vertices of~$u$ are not neighbors of~$v$,
then one end-vertex of the edges to these neighbors changes from~$u$ to~$v$,
and the position of these edges in the neighbors' edge lists
must be corrected to retain the sorting.

Let $n$ be the number of clusters (initially the vertex count), $m$
be the number of adjacent cluster pairs (edge count), and $d$ be the height
of the merge tree.  Merging the edge lists of two clusters has linear runtime
in the list lengths, and each edge participates in at most $d$ such merges.
Thus the worst-case runtime is $\bigO(d m)$ for the merges and,
given that the length of each edge list is at most~$n$,
$\bigO(d m n)$ for the position corrections.
(The implementation of Clauset et~al.~\cite{clauset2004vln}
has better worst-case bounds, but experimental results in Section~\ref{s:experiment}
indicate that it is not more efficient in practice.)

In order to quickly find the prioritized cluster pair for the next
merge, a priority queue (max-heap) over the clusters and their current
best partner is used.  It is updated as described
in~\cite{wakita2007msn}.  In worst case the priority queue is updated
with each merged edge, taking $\bigO(d m \log n)$ runtime.


\subsection{Multi-Step Greedy.}\label{ss:multistep}

To prevent extremely unbalanced cluster growth, Schuetz and Caflisch
introduced Multi-Step Greedy coarsening, which iteratively merges the
$l$~{\em disjoint} cluster pairs with the highest priority
(unless the merge decreases the modularity)~\cite{schuetz2008emo}.
Single-Step Greedy coarsening corresponds to the special case of $l=1$
(at least conceptually, the implementation differs).
To make the parameter~$l$ independent of the graph size, we specify it
as percentage of the number of modularity-increasing cluster pairs,
and call it \emph{merge fraction}.%
\footnote{Recently, Schuetz and Caflisch provided the empirical formula
  $l_{\mathop{opt}}\mathop{:=}\alpha\sqrt{f(V,V)}$ for good values of
  $l$~\cite{schuetz2008mga}.
  It does not outperform our formula for unweighted graphs
  (see Section~\ref{s:related}), and is unsuitable for weighted graphs,
  because scaling all edge weights with a positive constant
  changes $l_{\mathop{opt}}$ but not the optimal clustering.
}
The impact of the merge fraction on the effectiveness of Multi-Step Greedy
coarsening will be examined experimentally in Section~\ref{ss:ecoarsening}.

\begin{algorithm2e}
  \SetKwData{graph}{graph}
  \SetKwData{clustering}{clustering}
  \SetKwData{ranker}{merge prioritizer}
  \SetKwData{mf}{merge fraction}
  \SetKwData{merged}{merged}
  \SetKwData{ip}{ip}
  \SetKwData{bp}{bp}
  \SetKwData{source}{source}
  \SetKwData{target}{target}
  \SetKw{brk}{break}
  \SetKw{True}{true}
  \SetKw{False}{false}

  \Titleofalgo{~Multi-Step Greedy Coarsening}
  \KwIn{\graph, \ranker, \mf}
  \KwOut{\clustering}
  \BlankLine

  initialize \clustering with singleton clusters\;
  \While{$\exists$ {\rm cluster pair} $(C,D): \Delta Q_{C,D}>0$}{
    $l \leftarrow \mf \times \big| \{ (C,D)$ : $\Delta Q_{C,D}\!>\!0 \} \big|$\;
    mark all clusters as unmerged\;

    \vspace{0.5ex}
    \For{$\lceil l \rceil$ {\rm most prioritized pairs} $(C,D)$}{
      \If{\rm $C$ and $D$ are marked as unmerged}{
        merge clusters $C$ and $D$\;
        mark clusters $C$ and $D$ as merged\;
      }
    }
  }
\end{algorithm2e}

\paragraph{Implementation and Runtime.}

The same basic data structures as in Single-Step Greedy are used.
To iterate over the $l$~cluster pairs in priority order, the edges are
sorted once before entering the inner loop. This requires $\bigO(m \log m)$
time in worst case.  Alternative implementations optimized for very
small merge fractions could use partial sorting with $\bigO(m \log l)$
(but a larger constant factor).  With merge fraction $\alpha$ the
inner loop is repeated at least $n/\alpha$ times.
However, if only few disjoint cluster pairs exist, as in some power-law graphs,
up to $n$~iterations may be necessary.

\subsection{Merge Prioritizers.}\label{ss:priority}

A merge prioritizer assigns to each cluster pair $(C,D)$ a real number
called {\em merge priority}, and thereby determines the order in which
the coarsening algorithms merge cluster pairs.
Because the coarsening algorithms use only the order of the priorities,
two prioritizers can be considered as equivalent
if one can be transformed into the other by adding a constant
or multiplying with a positive constant.

The {\em Modularity Increase (MI)} $\Delta Q_{C,D}$ resulting
from the merge of the clusters $C$ and~$D$ is an obvious and widely used
merge prioritizer \cite{newman2004fad,clauset2004vln,schuetz2008emo,Ye2008aca}.

The {\em Weight Density (WD)} is defined as $\frac{f(C,D)}{\deg(C)\deg(D)}$,
and is equivalent to $\frac{\Delta Q_{C,D}}{\deg(C)\deg(D)}.$
Its use as merge prioritizer has not yet been proposed in the literature,
although Newman and Girvan originally introduced the modularity measure
to formalize the requirement of intra-cluster density
and inter-cluster sparsity~\cite{newman2004fae},
and Reichardt and Bornholdt showed that clusterings with optimal modularity
indeed fulfill this requirement~\cite{reichardt2006smc}.

The {\em Significance (Sig)}, another new merge prioritizer, is defined as
$\frac{\Delta Q_{C,D}}{\sqrt{\deg(C)\deg(D)}}$, and is thus
a natural compromise between Modularity Increase and Weight Density.
A further motivation is its relation to the (im)probability
of the edge weight~$f(C,D)$ in the null model described
in Section~\ref{ss:modularity}.
Under this null model, both the expected value and
the variance (at least for large $\deg(V)$) of the edge weight
between $C$ and~$D$ are $\frac{\deg(C)\deg(D)}{\deg(V)}$,
and the Significance is equivalent to the number of standard deviations
that separate the actual edge weight from the expected edge weight.

{\em Danon et al. (DA)}\ observed that the Modularity Increase $\Delta Q_{C,D}$
tends to prioritize pairs of clusters with large degrees, and proposed
the merge prioritizer $\frac{\Delta Q_{C,D}}{\min(\deg(C),\deg(D))}$
to avoid this bias~\cite{danon2006esh}.
It equals the Significance if $\deg(C) \mathop{=} \deg(D)$,
and is another compromise between Modularity Increase and Weight Density.

{\em Wakita and Tsurumi} found that greedy coarsening by Modularity Increase
tends to merge clusters of extremely uneven sizes~\cite{wakita2007msn}.
In order to suppress unbalanced merges, they proposed the merge prioritizer
$\min\big(\frac{\mathrm{size}(C)}{\mathrm{size}(D)},
          \frac{\mathrm{size}(D)}{\mathrm{size}(C)}\big) \Delta Q_{C,D}$,
where $\mathrm{size}(C)$ is either the number of vertices in $C$
(prioritizer {\em HN}) or the number of other clusters to which $C$ is connected
by an edge of positive weight (prioritizer {\em HE}).

Other types of merge prioritizers are clearly possible.
For example, vertex distances from random walks or eigenvectors
of certain matrices have been successfully applied in several clustering
algorithms (e.g., \cite{pons2006ccl,newman2006fcs,donetti2004dnc}).
However, preliminary experiments suggest that these relatively complicated
and computationally expensive prioritizers may not be more effective than
the simple prioritizers in this section~\cite{rotta2008mla}.

\vspace{-2mm}
\section{\label{s:refinement}Refinement Algorithms}

Refinement algorithms perform a local search by iteratively moving
individual vertices to different clusters (including newly created
clusters) such that the modularity increases.
\pagebreak

The first three subsections describe simple variants of greedy refinement,
and the final subsection proposes, for the first time in modularity clustering,
to apply refinement on more than one level of the coarsening hierarchy.
Excluded from consideration are algorithms with several tunable parameters
or explicit randomness, like simulated annealing
\cite{reichardt2006smc,medus2005dcs,massen2005icel}
or extremal optimization~\cite{duch2005cdeo}.

\subsection{Complete Greedy.}

Complete Greedy refinement repeatedly performs the best vertex move,
until no further modularity-increasing vertex moves are possible.
Here the \emph{best} vertex move is a move with the largest modularity
increase $\Delta Q_{v \to D}$ over all vertices~$v$ and all target
clusters~$D$.

\begin{algorithm2e}
  \SetKwData{clustering}{clustering}
  \SetKwData{graph}{graph}
  \SetKwData{CStart}{start}
  \SetKwData{CPeak}{peak}

  \Titleofalgo{~Complete Greedy Refinement}
  \KwIn{\graph, \clustering}
  \KwOut{\clustering}
  \BlankLine

  \Repeat{$\Delta Q_{v\to D} \leq 0$}{
    $(v,D) \leftarrow$ best vertex move\;
    \If{$\Delta Q_{v\to D}>0$}{
      move vertex $v$ to cluster $D$\;
    }
  }
\end{algorithm2e}

\paragraph{Implementation and Runtime.}

Vertices are moved in constant time using a vector mapping vertices to
their current cluster.  To find the best move, the modularity changes
$\Delta Q_{v\to D}$ for all vertices~$v$ and clusters~$D$ adjacent
to~$v$ (and a newly created cluster) need to be
determined.  For this purpose the algorithm iterates over the
vertices.  For each vertex~$v$ the summed weights $f(v,D)$ are
collected in one pass over its edges by using a search
tree similar to~\cite{blondel_fast_2008}.  Given $f(v,D)$,
the modularity change $\Delta Q_{v\to D}$ can be computed in constant time
(see Section~\ref{ss:modularity}).
Therefore, to find the globally best move,
all $n$~vertices and $m$~edges are visited once, and the weight of
each edge is added in a search tree of at most $n$~entries.  Assuming
$\bigO(n)$ moves yields a worst-case runtime of $\bigO(n m \log n)$.

\subsection{Fast Greedy.}

Fast Greedy refinement repeatedly iterates through all vertices
and moves each vertex to its best cluster,
until no improvement is found for any vertex.
Finding the best move for a particular vertex is considerable cheaper
than finding the globally best vertex move, as in Complete Greedy refinement;
the question whether this improved efficiency comes at the cost of worse
effectiveness will be addressed by an experiment in Section~\ref{ss:erefinement}.
Fast Greedy refinement has been previously proposed
by Schuetz and Caflisch~\cite{schuetz2008emo} and Ye et~al.~\cite{Ye2008aca}.

\begin{algorithm2e}
  \SetKwData{clustering}{clustering}
  \SetKwData{graph}{graph}
  \SetKwData{CStart}{start}
  \SetKwData{CPeak}{peak}

  \Titleofalgo{~Fast Greedy Refinement}
  \KwIn{\graph, \clustering}
  \KwOut{\clustering}
  \BlankLine

  \Repeat{{\rm no improved clustering found}}{
    \ForEach{{\rm vertex }$v$}{
      $D \leftarrow $ best cluster for $v$\;
      \If{$\Delta Q_{v\to D}>0$}{
        move vertex $v$ to cluster $D$\;
      }
    }
  }
  \vspace{-4mm}
\end{algorithm2e}

\paragraph{Implementation and Runtime.}

The implementation is very similar to the previous algorithm.
The worst-case time for one run of the inner loop is $\bigO(m \log n)$,
and a few runs usually suffice.

The order in which the inner loop visits the vertices
seems to have little impact on the obtained modularity in practice;
in our implementation, vertices are sorted by increasing number of edges.

\subsection{Adapted Kernighan-Lin.}

Kernighan-Lin refinement extends Complete Greedy refinement
with a basic capability to escape local maxima.
The algorithm was originally proposed by Kernighan and Lin
for minimum cut partitioning~\cite{kernighan1970ehp},
and was adapted to modularity clustering by Newman~\cite{newman2006mac}
(though with a limitation to two clusters).
In its inner loop, the algorithm
iteratively performs the best vertex move, with the restriction
that each vertex is moved only once, but without the restriction
that each move must increase the modularity.
After all vertices have been moved,
the inner loop is restarted from the best found clustering.
Preliminary experiments indicated that it is much more efficient
and rarely less effective to abort the inner loop
when the best found clustering has not improved
in the last $k:=10 \log_2|V|$ vertex moves~\cite{rotta2008mla}.

\begin{algorithm2e}
  \SetKwData{clustering}{clustering}
  \SetKwData{graph}{graph}
  \SetKwData{CStart}{start}
  \SetKwData{CPeak}{peak}
  \SetKw{brk}{break}

  \Titleofalgo{~Adapted Kernighan-Lin Refinement}
  \KwIn{\graph, \clustering}
  \KwOut{\clustering}
  \BlankLine

  \Repeat{{\rm no improved clustering found}}{
    $\CPeak \leftarrow \clustering$\;
    mark all vertices as unmoved\;

    \vspace{0.5ex}
    \While{{\rm unmoved vertices exist}}{
      $(v,D) \leftarrow $ best move with $v$ unmoved\;
      move $v$ to cluster $D$, mark $v$ as moved\;
      \If{$Q(\clustering) > Q(\CPeak)$}{
        $\CPeak \leftarrow \clustering$\;
      }
      \If{$k$ {\rm moves since last peak}}{\brk\;}
    }

    $\clustering \leftarrow \CPeak$\;
  }
\end{algorithm2e}

\paragraph{Implementation and Runtime.}

The implementation is largely straightforward; to improve efficiency,
the current clustering is not copied unless the modularity begins to decrease.
The worst case runtime is the same as for Complete Greedy,
assuming that a few outer iterations suffice.
In practice, Kernighan-Lin refinement takes somewhat longer, because it also
performs modularity-decreasing moves.

\subsection{Multi-Level Refinement.}
The refinement algorithms in the previous subsections easily get stuck
in suboptimal clusterings because they only move individual vertices.
Even Kernighan-Lin refinement is unlikely to move a medium-sized group
of densely interconnected vertices to another cluster, because
this would require a series of sharply modularity-decreasing vertex moves.
However, the vertex group may well have been merged into a single cluster
at some stage of the coarsening, and a refinement algorithm can easily
reassign the group if it moves entire clusters of this coarsening level,
instead of individual vertices.
This is the basic idea of multi-level refinement, which has already proved
to be very effective for minimum cut partitioning problems
\cite{hendrickson1995mlp,karypis1998fah}.

The Multi-Level Clustering algorithm first executes a coarsening algorithm
(for example, any algorithm from Section~\ref{s:coarsening})
and then, usually several times, a refinement algorithm
(for example, any algorithm from the previous subsections).
Intermediate results of the coarsening algorithm are recorded
as {\em coarsening levels} whenever the number of clusters has decreased
by a certain percentage, which is provided as a parameter
called {\em reduction factor}.
Each coarsening level is a graph whose vertices are the clusters
at the respective state of coarsening.
The refinement algorithm is applied to every coarsening level,
from the coarsest level to the original graph.
At the coarsest level, each vertex belongs to a separate cluster,
and at the finer levels, the initial cluster membership of each vertex
is copied from the corresponding vertex of the previous (coarser) level.

\begin{algorithm2e}
  \SetKwData{graph}{graph}
  \SetKwData{level}{level}
  \SetKwData{clustering}{clustering}
  \SetKwData{rf}{reduction factor}
  \SetKwFunction{coarsener}{coarsener}
  \SetKwFunction{refiner}{refiner}
  \SetKw{brk}{break}
  \SetKw{KwFrom}{from}

  \Titleofalgo{~Multi-Level Clustering}
  \KwIn{\graph, \coarsener, \refiner, \rf}
  \KwOut{\clustering}
  \BlankLine

  \tcp{coarsening phase}
  $\level[1] \leftarrow \graph$\;
  \For{$l$ \KwFrom 1 \KwTo \ldots}{
    $\level[l\!+\!1] \leftarrow$ \coarsener{$\level[l]$, \rf}\;
    \lIf{{\rm no clusters merged}}{\brk\;}
  }

  \vspace{0.5ex}
  \tcp{refinement phase}
  $\clustering \leftarrow$ vertices of $\level[l_{\max}]$\;
  \For{$l$ \KwFrom $l_{\max}-1$ \KwTo 1}{
    project \clustering from $\level[l\!+\!1]$ to $\level[l]$\;
    $\clustering \leftarrow $\refiner{$\level[l]$, \clustering}\;
  }
\end{algorithm2e}

The conventional Single-Level refinement, which executes
a refinement algorithm only on the original graph,
is the special case of Multi-Level refinement with a reduction factor of 100\%.
While Multi-Level refinement, and more generally every decrease
of the reduction factor, potentially produces better clusterings,
it may be suspected to significantly increase the required runtime.
However, this is not necessarily the case, because the additional coarsening
levels are smaller than the original graph, and one cheap vertex move
on a coarse graph can save many expensive vertex moves on a finer graph.
The impact of the reduction factor on both effectiveness and efficiency
is examined experimentally in Section~\ref{ss:erefinement}.

\paragraph{Related Work.}

Several recent algorithms for modularity clustering are related
to Multi-Level refinement, but differ in crucial respects.
Djidjev's method is (despite its name) not itself a multi-level algorithm,
but a divisive method built on an existing multi-level algorithm
for minimum cut partitioning~\cite{djidjev2006csd}.
Blondel et~al.\ use local search on multiple levels to coarsen graphs,
but do not refine the results of the coarsening~\cite{blondel_fast_2008}.
Ye et~al.'s algorithm performs refinement on multiple coarsening levels,
but only moves vertices of the original graph instead
of coarse vertices (clusters)~\cite{Ye2008aca}.

\paragraph{Implementation and Runtime.}

With reduction factor $\alpha$ at most $\log_{1/(1-\alpha)}(n)$
coarsening levels are generated.  For each new level a graph
homomorphism connecting it to the previous level is constructed and
used to transfer weights and clusterings between levels.  To decouple
the Multi-Level Clustering algorithm from details of the coarsening algorithm,
the coarse graph and its homomorphism is constructed from the clustering
produced by the coarsening algorithm: The vertices of each cluster are
connected to their cluster-vertex in $\bigO(n)$ time.  For each edge
the corresponding cluster-edge has to be found or added
if not yet existing.  This search is accelerated by processing all
vertices of a cluster successively and using a search tree over
the cluster-edges of the current cluster.  Thus
constructing and connecting all edges costs $\bigO(m \log n)$ time.

\section{\label{s:experiment}Experiments}

This section experimentally compares the effectiveness (achieved
modularity) and efficiency (runtime) of the various heuristics presented in
the previous sections.

\subsection{Experimental Setup.}

The heuristics were implemented in C++ and compiled with GCC~$4.2.3$.
The implementations are available online at
\url{http://www.informatik.tu-cottbus.de/~rrotta/}.

In order to compare the effectiveness of the heuristics, the
arithmetic mean of the modularity over a fixed set of graphs is measured;
higher means indicate more effective algorithms.
(Thus only the relative values of the means are interpreted,
the absolute values are not intended to be meaningful.)
Generated graphs are not used because they are
structurally very limited (e.g., in their vertex degree distribution), and
do not necessarily permit generalizations to graphs from real applications.
Instead the graph set contains 58 real-world graphs retrieved from various
resources as listed in Appendix~\ref{app:graphs}.  The available graphs were
roughly classified by their application domain and
graphs of diverse size that fairly represent all major domains were selected.
In addition the collection includes commonly used benchmark
graphs like Zachary's karate club network~\cite{zachary1977ifm}.  The
graphs range from a few to 75k vertices and 352k edges.

All runtimes were measured on a 3.00GHz Intel Pentium 4 processor
with 1GB main memory.
The time for reading the graph was excluded to avoid that
it interferes with the aspects studied here.

\subsection{\label{ss:ecoarsening}Coarsening Algorithms.}

\begin{figure}
  \includegraphics[width=\columnwidth]{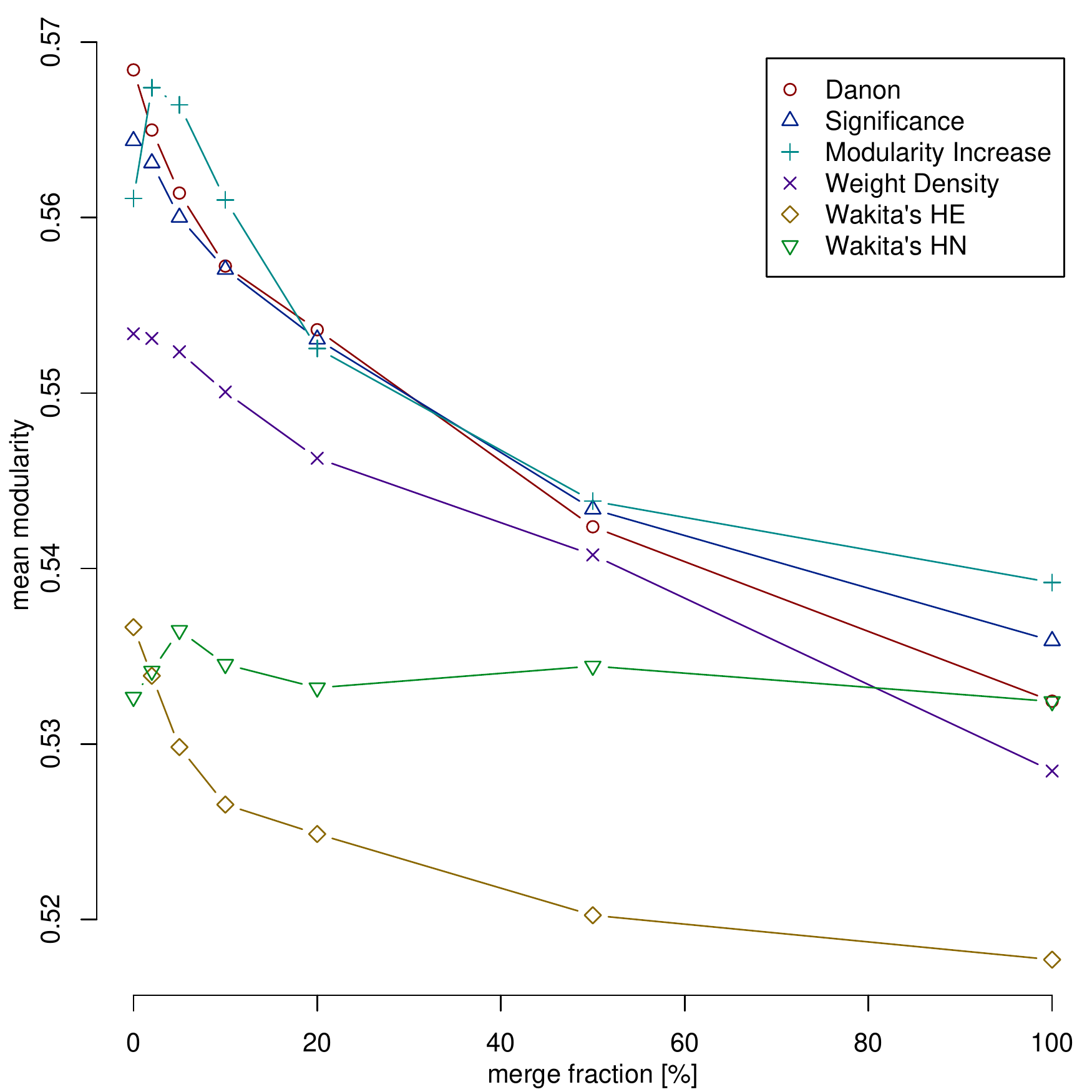}
  \caption{Modularity by merge fraction and prioritizer.}
  \label{fig:selector-mf}
\end{figure}

\begin{figure}
  \includegraphics[width=\columnwidth]{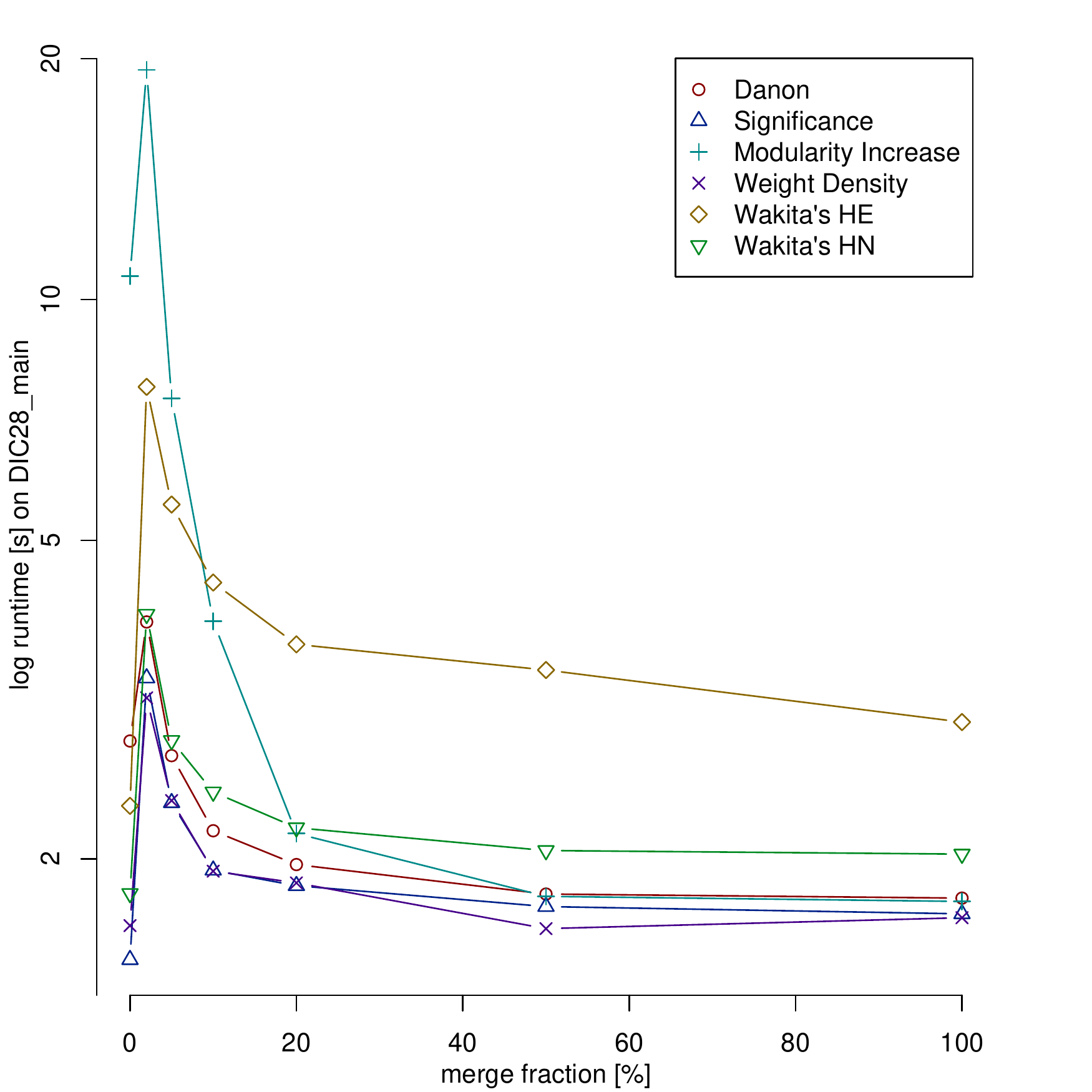}
  \caption{Runtime by merge fraction and prioritizer on the graph
    `DIC28\_main'.}
  \label{fig:selector-mf-runtime}
\end{figure}

Figure~\ref{fig:selector-mf} compares the effectiveness of the merge prioritizers
for Single-Step Greedy coarsening (represented by a merge fraction of~0\%)
and Multi-Step Greedy coarsening with merge fractions of 2\%, 5\%,
10\%, 20\%, 50\%, and 100\%.  No refinement was used.  The runtime
measured on the graph `DIC28\_main' is shown in
Fig.~\ref{fig:selector-mf-runtime} and is typical for larger
graphs.

Concerning the merge prioritizers, Wakita's HE and~HN are much less effective
than the others, and not more (usually even less) efficient.
The lower effectiveness is also visible in Fig.~\ref{fig:extalg-mod-ml}
for Wakita's original implementation.

Concerning the algorithms, Multi-Step Greedy is generally less
effective and less efficient than the simpler Single-Step Greedy.
Only for Modularity Increase, Multi-Step Greedy is faster
and, for merge fractions of 2\% and~5\%, also slightly more effective,
but still similar to Single-Step Greedy with Danon and Significance.
Apparently the other merge prioritizers do not benefit
from Multi-Step Greedy's tendency to balance cluster sizes
because, unlike Modularity Increase, they have no strong bias
towards merging large clusters.

\begin{figure}
  \includegraphics[width=\columnwidth]{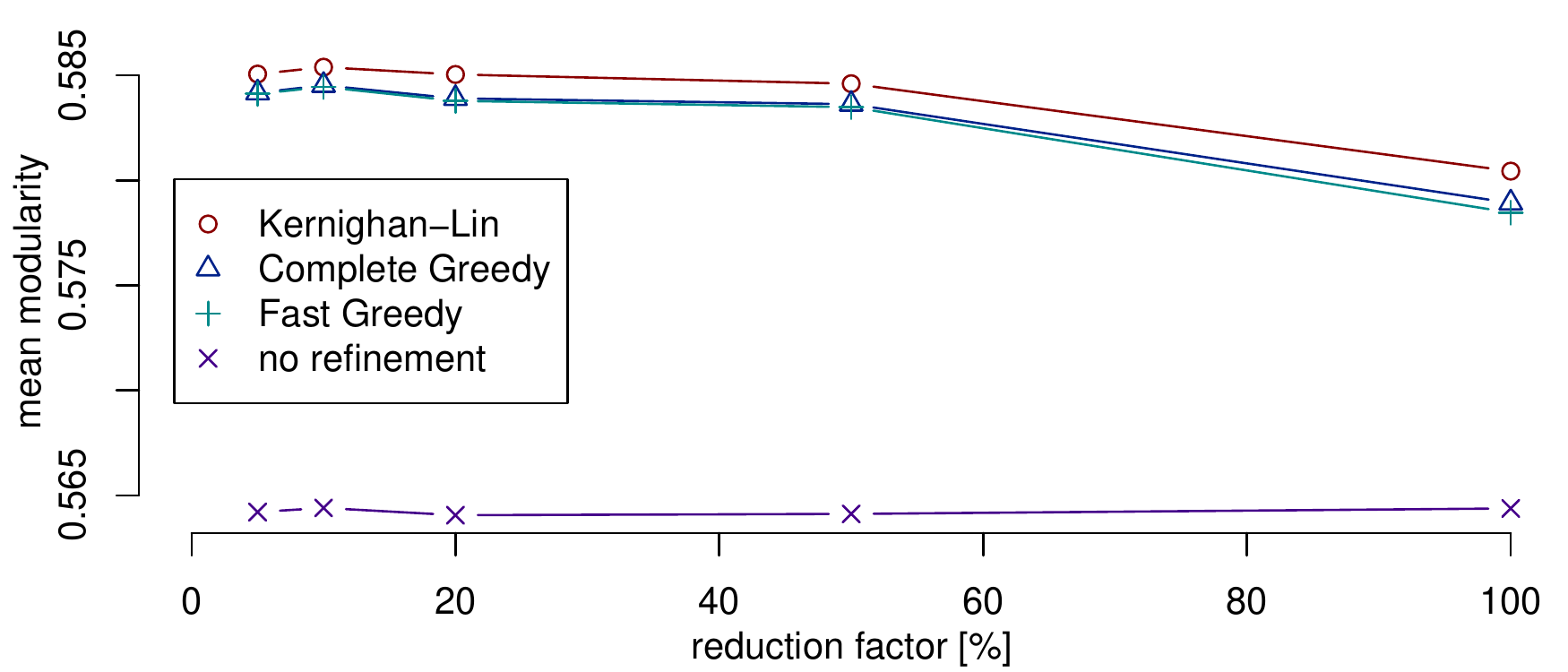}
  \caption{Modularity by reduction factor and refinement method.}
  \label{fig:reduction-meanmod}
\end{figure}

\begin{figure}
  \includegraphics[width=\columnwidth]{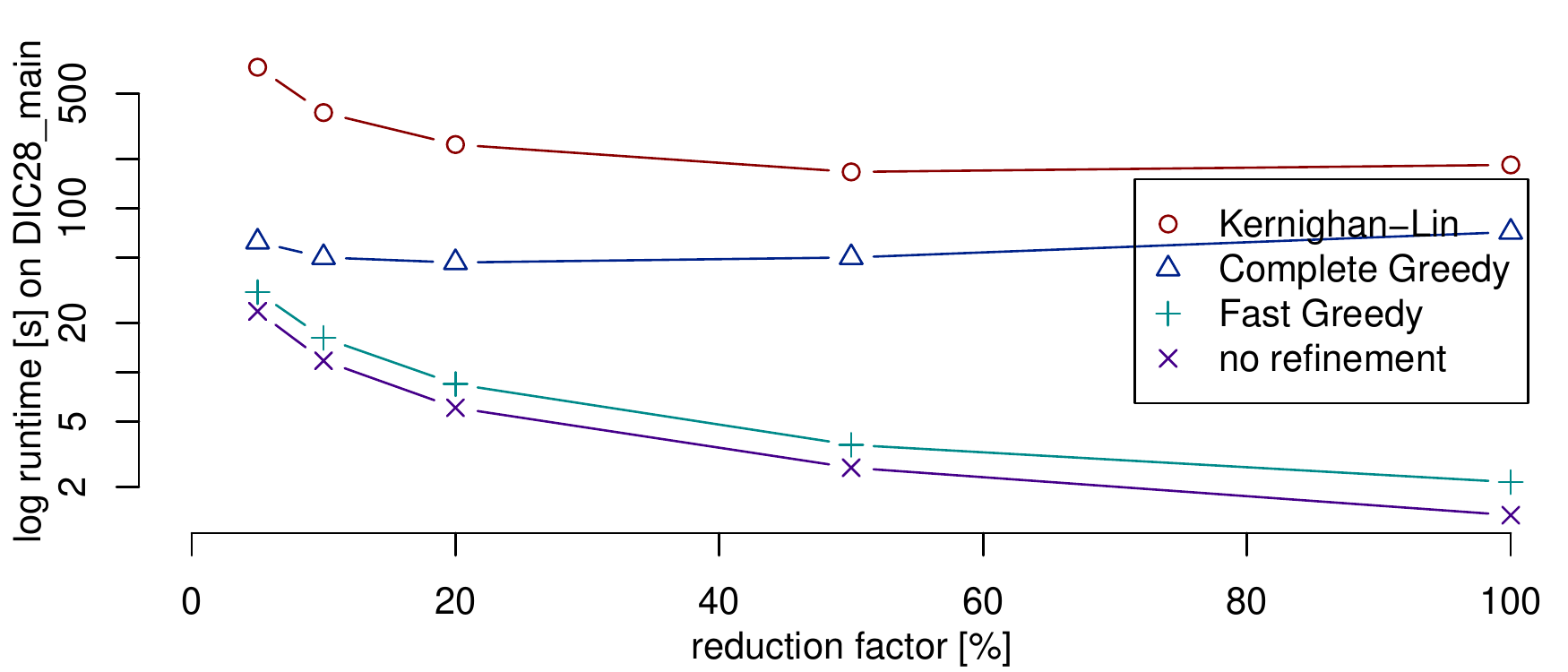}
  \caption{Runtime by reduction factor and refinement method on the graph `DIC28\_main'.}
  \label{fig:reduction-runtime-dic28}
\end{figure}

\begin{figure*}
  \centering
  \includegraphics[width=\textwidth]{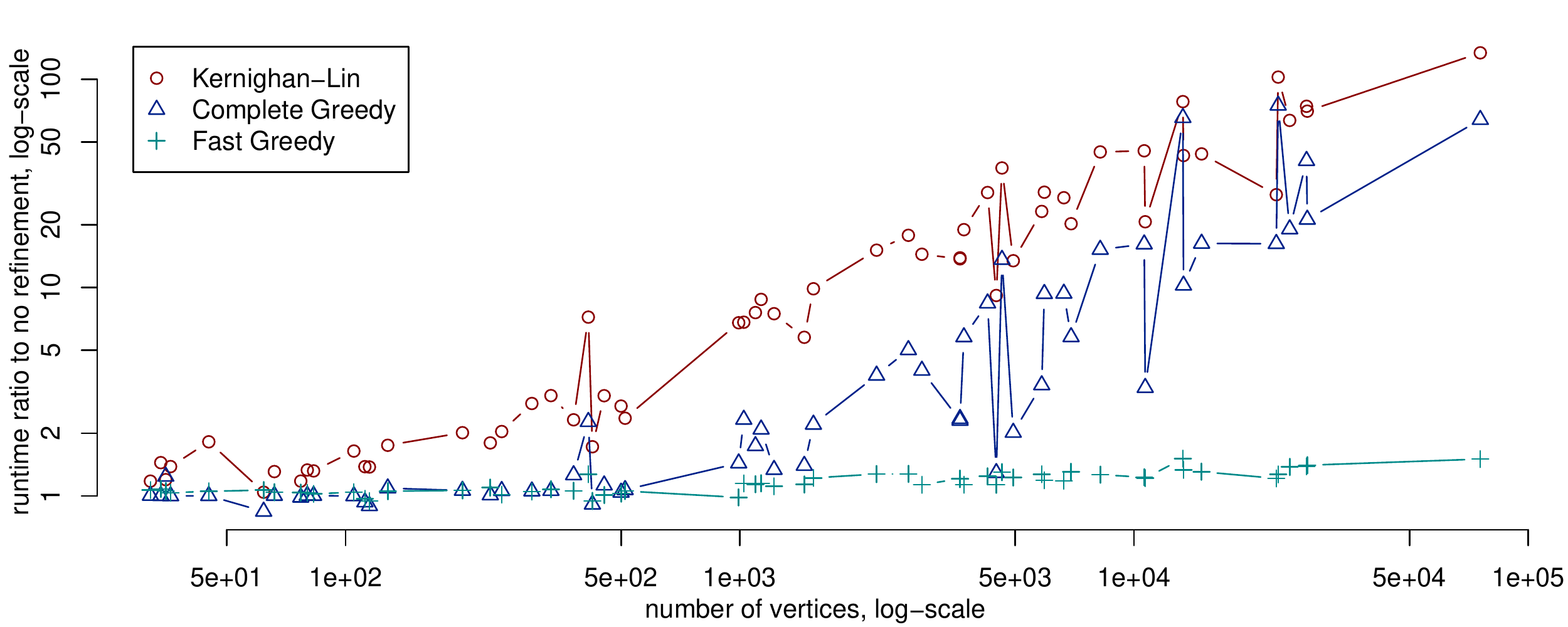}
  \caption{Runtime ratio of the refinement heuristics to raw
    coarsening, log-log scaled. Reduction factor is 50\%.}
  \label{fig:reduction-runtime}
\end{figure*}

\subsection{\label{ss:erefinement}Refinement Algorithms.}

Figure~\ref{fig:reduction-meanmod} compares the effectiveness
of the refinement algorithms
for Single-Level refinement (reduction factor 100\%)
and Multi-Level refinement with reduction factors of 5\%,
10\%, 20\%, and~50\%. As coarsener Single-Step Greedy with the Significance
prioritizer was chosen, because it proved to be effective and efficient
in the previous subsection.  The runtime measurements on `DIC28\_main'
are shown in Fig.~\ref{fig:reduction-runtime-dic28} and the
dependency of the runtime on the graph size is depicted in
Fig.~\ref{fig:reduction-runtime}.

Multi-Level refinement with a reduction factor of~50\% turns out to be
more effective than Single-Level refinement, and similarly efficient.
Reduction factors below 50\% do not considerably improve the modularity,
but significantly increase the runtime for Fast Greedy.

Fast Greedy refinement is about as effective as Complete Greedy, and
just slightly less effective than Kernighan-Lin, but much faster. It
scales well with the graph size (see Fig.~\ref{fig:reduction-runtime}),
while Complete Greedy and Kernighan-Lin become prohibitively expensive.

\subsection{\label{ss:ecombi}Combining Coarsening and Refinement.}

\begin{figure}
  \includegraphics[width=\columnwidth]{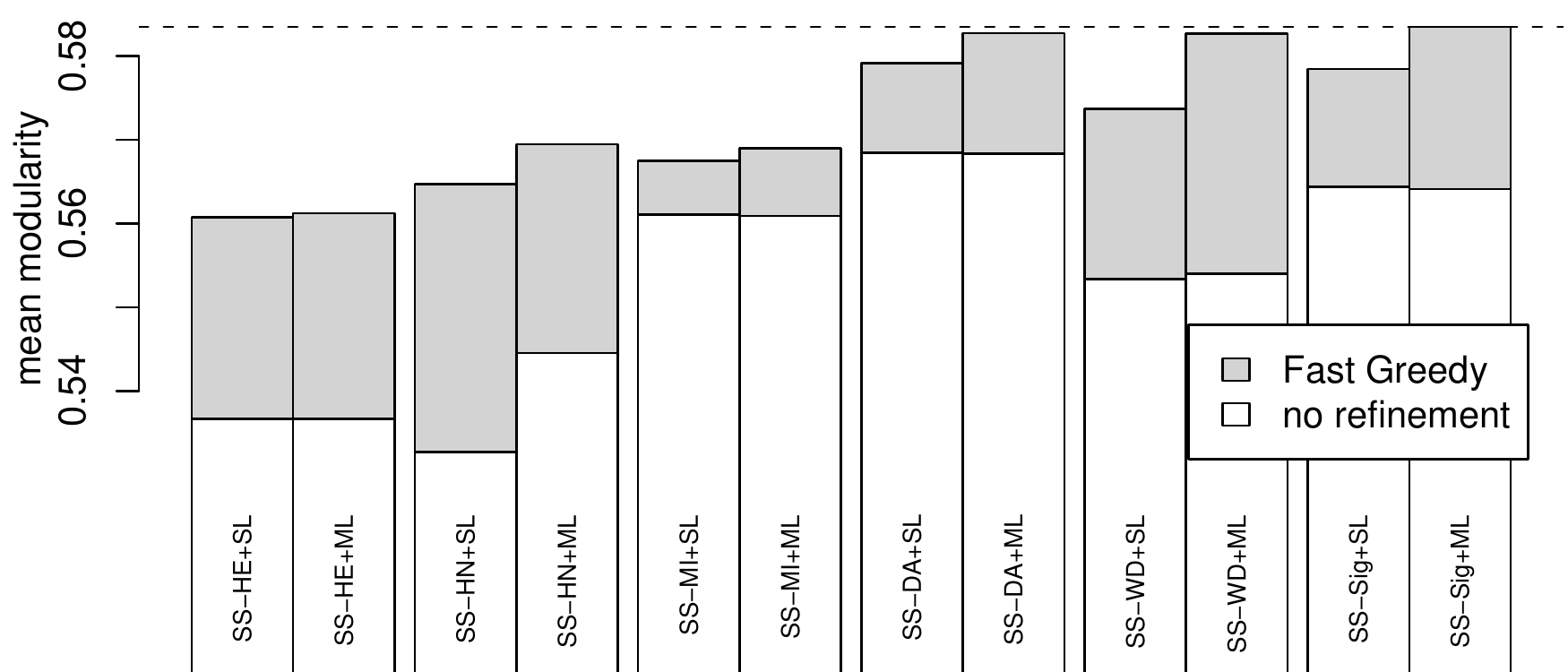}
  \caption{Mean modularity by merge prioritizer. Left bars show
    reduction factor 100\% (Single-Level), right bars 50\%
    (Multi-Level). Both use Single-Step Greedy.}
  \label{fig:selector2}
\end{figure}

Concerning Single-Level vs.\ Multi-Level refinement,
Fig.~\ref{fig:selector2} shows that Multi-Level refinement
is consistently more effective for all merge prioritizers,
and thus confirms the results for the Significance prioritizer
in Fig.~\ref{fig:reduction-meanmod}.

\begin{figure}
  \includegraphics[width=\columnwidth]{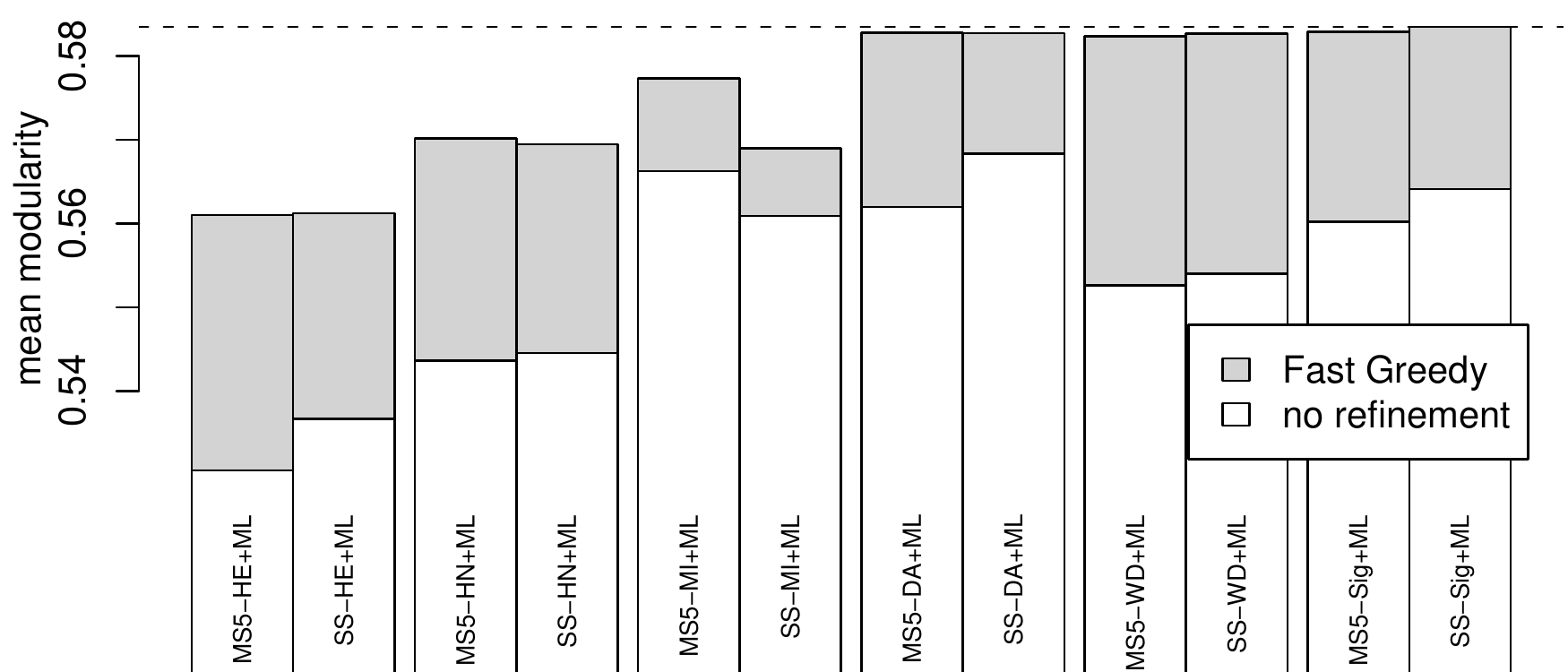}
  \caption{Mean modularity by merge prioritizer: Left bars show merge
    fraction 5\% (Multi-Step) and right bars 0\%
    (Single-Step). Both use reduction factor 50\%.}
  \label{fig:selector3}
\end{figure}

Concerning Single-Step vs.\ Multi-Step Greedy coarsening,
Fig.~\ref{fig:selector3} shows that for the best merge prioritizers,
both are similarly effective with Multi-Level refinement,
while Single-Step Greedy is more effective without refinement.
Clearly, Multi-Level refinement benefits from the uniform cluster growth
enforced by Multi-Step Greedy coarsening.
Overall, Single-Step Greedy coarsening is still preferable
because of its greater simplicity and efficiency.

Concerning the merge prioritizers, Figs.~\ref{fig:selector2}
and~\ref{fig:selector3} show that Modularity Increase is only competitive
without refinement (ignoring efficiency), and Weight Density is only competitive
with Multi-Level refinement.
Here Multi-Level refinement benefits from the bias of Weight Density
towards balanced cluster growth, and suffers from the bias of Modularity Increase
towards unbalanced cluster growth.
Danon and Significance are effective with and without refinement.

\subsection{Conclusions.}

The best algorithm found in these experiments
is Single-Step Greedy coarsening
with Danon or Significance as merge prioritizer
combined with Multi-Level Fast Greedy refinement
(or Multi-Level Kernighan-Lin, if efficiency is no concern).

Interestingly, Single-Step Greedy refinement outperformed
the recent and more complex Multi-Step Greedy (for the best merge prioritizers),
the Danon and Significance merge prioritizers clearly outperformed
the much more widely used Modularity Increase (especially with refinement,
and considering efficiency) and Wakita's prioritizers,
and the newly proposed Multi-Level refinement consistently outperformed
the popular Single-Level refinement.

\section{\label{s:related}Related Algorithms}

An exhaustive review and comparison of the numerous algorithms
for modularity clustering is beyond the scope of this paper;
the purpose of this section is to provide evidence that our recommended heuristic
-- Single-Step Greedy coarsening by Significance
with Multi-Level Fast Greedy refinement ({\em SS+ML}) --
is competitive with the best existing methods.

\subsection{Basic Approaches.}

Algorithms for modularity clustering can be categorized
into the following four types: \emph{Subdivision} heuristics try
to divide the network,
for example by iteratively removing edges~\cite{newman2004fae}
or by recursively splitting the graph using eigenvectors~\cite{newman2006mac}.
\emph{Coarsening} (or agglomeration)
heuristics iteratively merge clusters starting from singletons.
Cluster pairs can be selected based on random walks
\cite{pons2006ccl,pujol2006dcs},
increase of modularity \cite{clauset2004vln,schuetz2008emo,Ye2008aca},
or other criteria \cite{wakita2007msn,danon2006esh,donetti2004dnc}.
\emph{Local search} heuristics move vertices between clusters,
with Kernighan-Lin-style and greedy search being the most prominent examples.
Other approaches include Tabu Search~\cite{arenas2008ascn},
Extremal Optimization~\cite{duch2005cdeo},
and Simulated Annealing~\cite{reichardt2006smc,medus2005dcs,massen2005icel}.
Finally, \emph{mathematical programming} approaches model
modularity maximization as a linear or quadratic programming problem
which can be solved with existing software packages
\cite{agarwal2008mmg,BrandesEtAl:TKDE2008,XuEtAl:EPJ2007}.

\subsection{Published Modularity Values.}

Table~\ref{tab:pub_results} compares modularity values from various publications
with the results of our heuristic~SS+ML.
%
Mathematical programming approaches consistently find better clusterings
than SS+ML, though by a very small margin;
however, they are computationally much more expensive
and do not scale to large graphs~\cite{agarwal2008mmg,XuEtAl:EPJ2007}.
Compared to the best algorithms in the three other classes, the results
of SS+ML are very competitive, and for large graphs significantly better.

\begin{table}
  \vspace{11mm}
  \centering
  \resizebox{\columnwidth}{!}{
\begin{tabular}{l|rl@{~~}ll@{~~}ll@{~~}ll@{~~}ll}
\hline
  graph & size & \multicolumn{2}{c}{subdivision} &
                 \multicolumn{2}{c}{coarsening} &
                 \multicolumn{2}{c}{local\;search} &
                 \multicolumn{2}{c}{math\;prog} &
                 SS+ML \\
\hline
  karate~\cite{zachary1977ifm} &
    34 &
    \cite{newman2006mac} & .419 &
    \cite{Ye2008aca} & .4198 &
    \cite{duch2005cdeo} & .4188 &
    \cite{agarwal2008mmg} & .4197 &
    .4197 \\
  dolphins~\cite{lusseau2003dolphins} &
    62 &
    \cite{newman2006mac} & \emph{.4893} &
    \cite{pons2006ccl} & \emph{.5171} &
    \cite{reichardt2006smc} & \emph{.5285} &
    \cite{XuEtAl:EPJ2007} & .5285 &
    .5276 \\
  polBooks~\cite{krebs_polbooks} &
    105 &
    \cite{newman2006mac} & \emph{.3992} &
    \cite{schuetz2008mga} & \emph{.5269} &
    \cite{blondel_fast_2008} & \emph{.5204} &
    \cite{agarwal2008mmg} & .5272 &
    .5269 \\
  afootball~\cite{girvan2002css} &
    115 &
    \cite{white2005sca} & .602 &
    \cite{Ye2008aca} & .605 &
    \cite{blondel_fast_2008} & \emph{.6045} &
    \cite{agarwal2008mmg} & .6046 &
    .6002 \\
  jazz~\cite{gleiser2003csj} &
    198 &
    \cite{newman2006mac} & .442 &
    \cite{danon2006esh} & .4409 &
    \cite{duch2005cdeo} & .4452 &
    \cite{agarwal2008mmg} & .445 &
    .4446 \\
  celeg\_metab~\cite{duch2005cdeo} &
    453 &
    \cite{newman2006mac} & .435 &
    \cite{schuetz2008emo} & .450 &
    \cite{duch2005cdeo} & .4342 &
    \cite{agarwal2008mmg} & .450 &
    .4452 \\
  email~\cite{guimer2003sscs} &
    1133 &
    \cite{newman2006mac} & .572 &
    \cite{danon2006esh} & .5569 &
    \cite{duch2005cdeo} & .5738 &
    \cite{agarwal2008mmg} & .579 &
    .5774 \\
  Erdos02~\cite{erdos_project} &
    6927 &
    \cite{newman2006mac} & \emph{.5969} &
    \cite{pujol2006dcs} & .6817 &
    \cite{reichardt2006smc} & \emph{.7094} &
     &  &
    .7162 \\
  PGP\_main~\cite{boguna2004msn} &
    11k &
    \cite{newman2006mac} & .855 &
    \cite{danon2006esh} & .7462 &
    \cite{duch2005cdeo} & .8459 &
     &  &
    .8841 \\
  cmat03\_main~\cite{newman2001sscn} &
    28k &
    \cite{newman2006mac} & .723 &
    \cite{Ye2008aca} & .761 &
    \cite{duch2005cdeo} & .6790 &
     &  &
    .8146 \\
  ND\_edu~\cite{albert1999idw} &
    325k &
     &  &
    \cite{clauset2004vln} & .927 &
    \cite{blondel_fast_2008} & .935 &
     &  &
    .9509 \\
  \hline
\end{tabular}
}
  \caption{Best published modularity values for four algorithm classes,
           compared to the modularity values for our heuristic SS+ML.
           Where possible, missing values were substituted with results
           from published implementations (shown in italics).}
  \label{tab:pub_results}
\end{table}

\subsection{Published Implementations.}

In order to directly compare our heuristics with existing algorithms,
a range of publicly available implementations was retrieved
from authors' websites and through the \emph{igraph} library of Cs{\'a}rdi
and Nepusz~\cite{igraph2006}.  Only a subset of the graph collection
could be used as some implementations cannot process graphs with weighted edges
or self-edges.  The employed 23~graphs range from a few to 75k
vertices and are marked with \emph{UW} in Appendix~\ref{app:graphs}.
In some of these graphs negligible differences in edge weights
and small amounts of self-edges were removed.

\begin{figure}
  \includegraphics[width=\columnwidth]{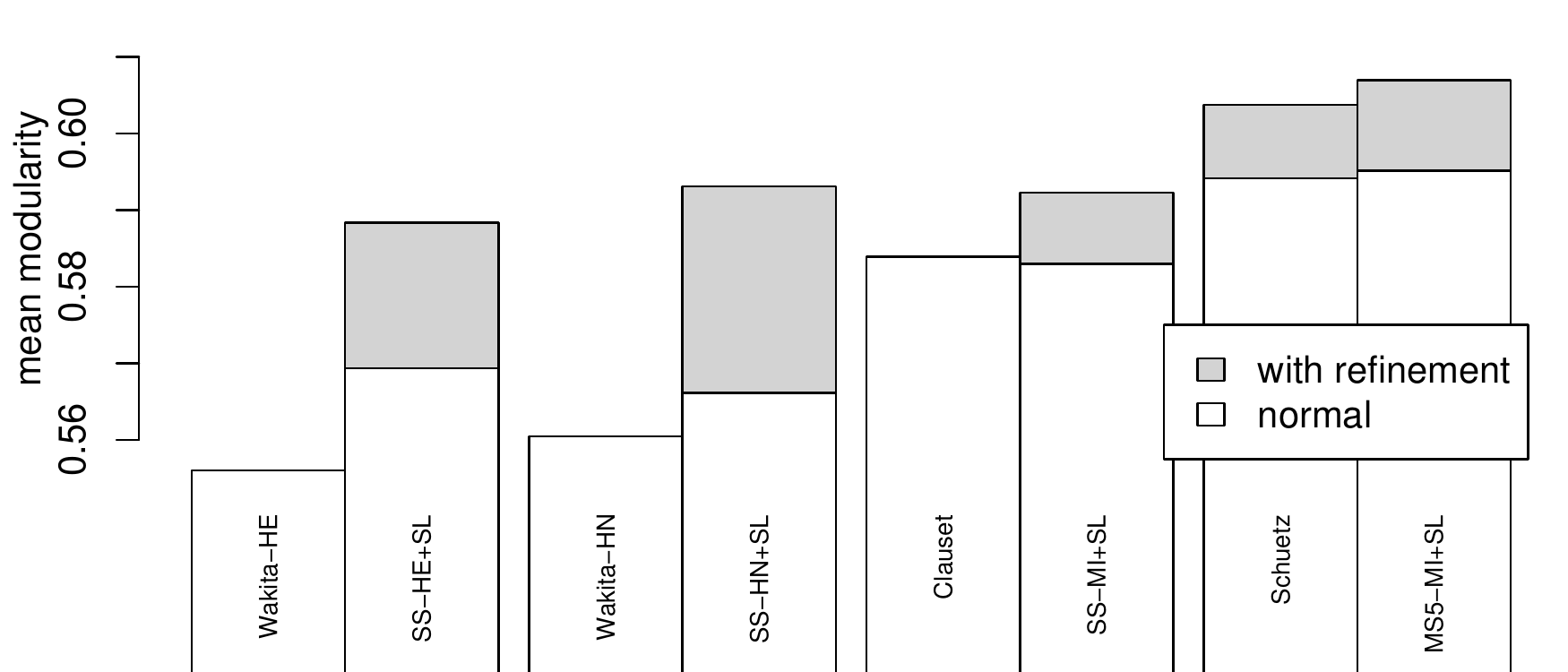}
  \caption{Mean modularities from four published implementations and
    our (approximate) reimplementations on unweighted graphs.}
  \label{fig:extalg-mod-eqv}
\end{figure}

\begin{figure}
  \includegraphics[width=\columnwidth]{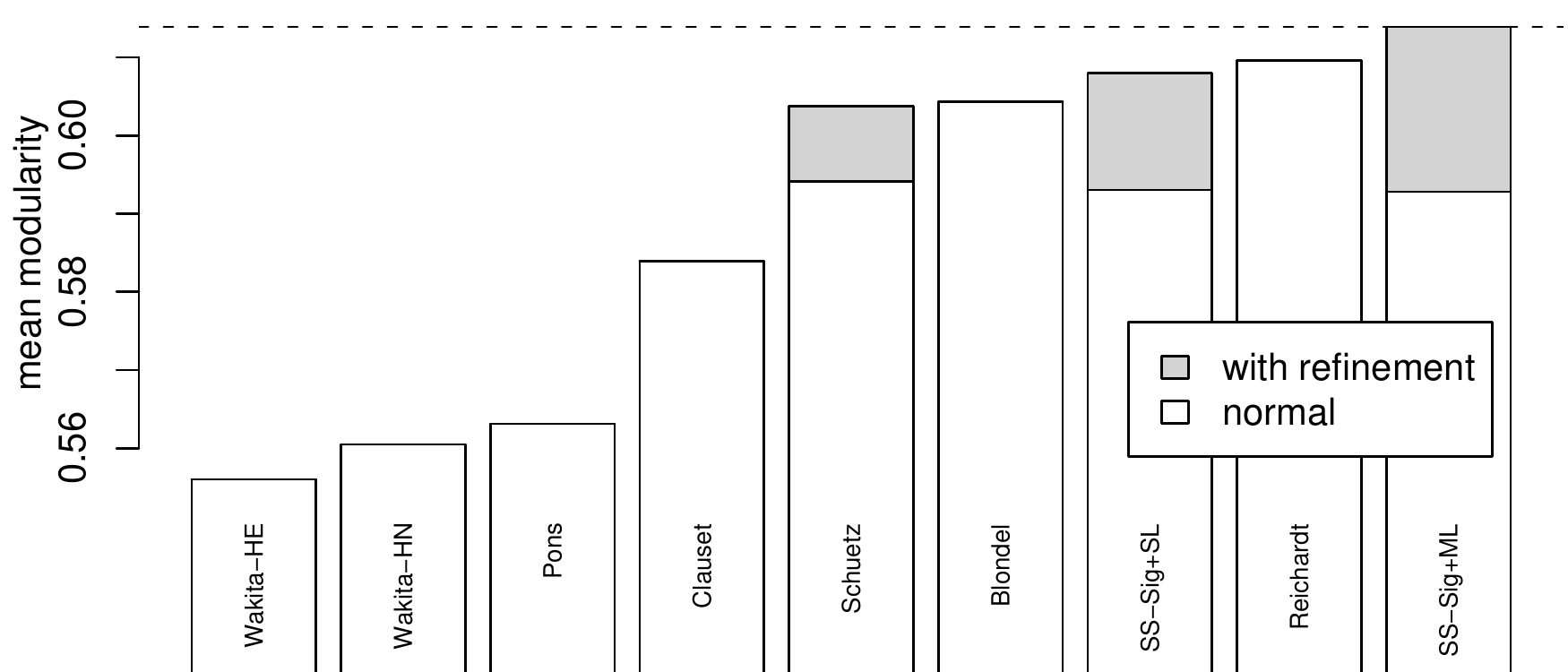}
  \caption{Mean modularities from the published implementations 
    and our recommended heuristic \emph{SS-Sig+ML} on unweighted graphs.}
  \label{fig:extalg-mod-ml}
\end{figure}

\begin{figure*}
  \centering
  \includegraphics[width=\textwidth]{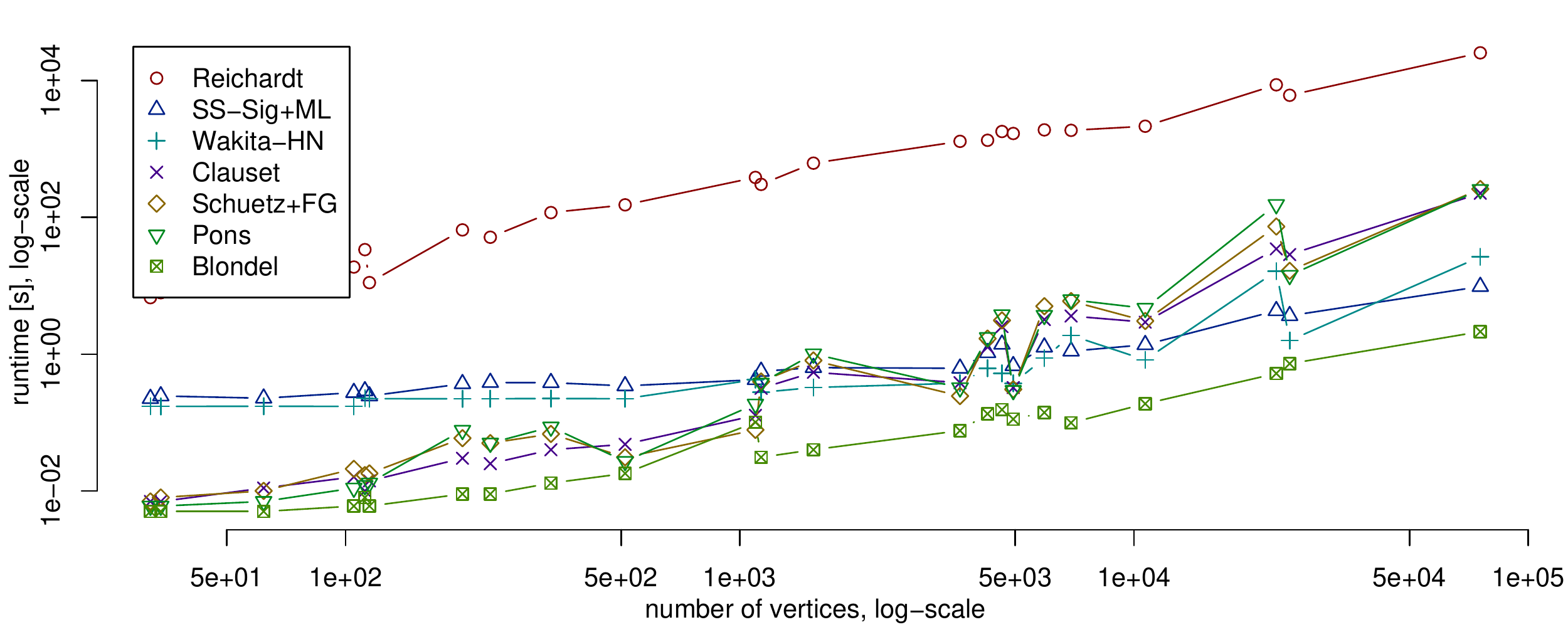}
  \caption{Runtime of the published implementations on unweighted graphs,
           log-log scaled.}
  \label{fig:extalg-runtime}
\end{figure*}

The included coarsening heuristics are the fast greedy joining
of Clauset et al.~\cite{clauset2004vln},
the algorithms of Wakita and Tsurumi~\cite{wakita2007msn},
the recent multi-step greedy algorithm
of Schuetz and Caflisch~\cite{schuetz2008emo}
(with parameter $l\mathop{=}0.25\sqrt{f(V,V)/2}$,
as recommended by Schuetz and Caflisch in~\cite{schuetz2008mga}),
and the algorithm of Pons and Latapy~\cite{pons2006ccl}
based on short random walks (here of length~$4$).
%
%
The examined local search heuristics are simulated annealing of Reichardt
and Bornholdt~\cite{reichardt2006smc} (here with at most 120 clusters)
and the recent hierarchical algorithm
of Blondel et al.~\cite{blondel_fast_2008}.

Concerning the performance of the (approximately) reimplemented heuristics,
the mean modularities from the published implementations 
are roughly reproduced or slightly improved 
by our implementations (Fig.~\ref{fig:extalg-mod-eqv}) -- 
even for Schuetz and Caflisch, where the computation  
of the parameter~$l$ differs (see Section~\ref{ss:multistep}).
Note that refinement is not available in the implementations 
of Wakita and Tsurumi and of Clauset et al., 
and is optional in the implementation of Schuetz and Caflisch.

Concerning the performance of our recommended heuristic,
Single-Step Greedy coarsening by Significance
with Multi-Level Fast Greedy refinement (\emph{SS-Sig+ML}),
only Reichardt and Bornholdt's implementation produces clusterings 
of similarly high modularity, but it is much slower, and only Blondel et al.'s
implementation is faster, but it produces worse clusterings
(Figs.~\ref{fig:extalg-mod-ml} and~\ref{fig:extalg-runtime}).
Even the still simpler and faster variant with Single-Level refinement 
(\emph{SS-Sig+SL}) produces competitive clusterings, notably in comparison
with the recent algorithm of Schuetz and Caflisch which is more complex 
and requires parameter tuning (see Section~\ref{ss:multistep}).


\section{Summary and Conclusion}

Various coarsening and refinement heuristics for modularity clustering
can be organized into a design space with four dimensions: merge fraction
(including Single-Step and Multi-Step Greedy coarsening), merge prioritizer,
refinement algorithm, and reduction factor (including Single-Level and
Multi-Level refinement).
In an experimental comparison of achieved modularities and runtimes,
some widely used or rather complex design alternatives --
for example, Multi-Step Greedy coarsening, merge prioritization
by Modularity Increase, or Single-Level refinement --
were outperformed by newly proposed or simpler alternatives --
particularly Single-Step Greedy coarsening by Significance
with Multi-Level Fast Greedy refinement.
In a comparison with published implementations and benchmark results,
this heuristic was more efficient than algorithms that achieved
similar modularities, and achieved higher modularities than algorithms
with similar or better efficiency.


\newpage
\appendix

\section{\label{app:graphs}The Benchmark Graph Collection}

Table~\ref{tab:graphs} on the next page lists graphs used for the
experiments.  The graphs postfixed with `\_main' just contain the
largest connectivity component of the original graph.  All graphs from
the subset `UW' were used without edge weights and self-edges for the
experiments on published implementations.  For each graph the source
collection is named in the last column.  Web addresses to these
collections are listed in Table~\ref{tab:sources}.  For information
about the original authors please visit the respective websites.

\begin{table*}
  \centering
  {\scriptsize 
\begin{tabular}{l|rrrrrr}
\hline
  & subset & vertices & edges & edge weight & type & source  \\
\hline
 SouthernWomen  & UW  &  32 &  89 &         89.0 &  social &  \href{http://vlado.fmf.uni-lj.si/pub/networks/data/GBM/default.htm}{pajek} \\
 karate  & UW   &  34 &  78 &         78.0 &  social &  \href{http://www-personal.umich.edu/~mejn/netdata/}{Newman} \\
 football  &    &  35 &  118 &        295.0 &  economy &  \href{http://vlado.fmf.uni-lj.si/pub/networks/data/sport/football.htm}{pajek} \\
 morse  &    &  36 &  666 &      25448.0 &  similarity &  \href{http://www.informatik.tu-cottbus.de/~an/GD/}{ANoack} \\
 Food  &    &  45 &  990 &      11426.0 &  similarity &  \href{http://www-sst.informatik.tu-cottbus.de/~an/GD/}{ANoack} \\
 dolphins  & UW   &  62 &  159 &        159.0 &  social &  \href{http://vlado.fmf.uni-lj.si/pub/networks/data/mix/mixed.htm}{pajek} \\
 WorldImport1999  &    &  66 &  2145 &    4367930.4 &  economy &  \href{http://www-sst.informatik.tu-cottbus.de/~an/GD/}{ANoack} \\
 lesmis  &    &  77 &  254 &        820.0 &  social &  \href{http://www-personal.umich.edu/~mejn/netdata/}{Newman} \\
 world\_trade  &    &  80 &  875 &   65761594.0 &  economy &  \href{http://vlado.fmf.uni-lj.si/pub/networks/data/esna/metalWT.htm}{pajek} \\
 A00\_main  &    &  83 &  135 &        135.0 &  software &  \href{http://vlado.fmf.uni-lj.si/pub/networks/data/GD/GD.htm}{GraphDrawing} \\
 polBooks  & UW   &  105 &  441 &        441.0 &  similarity &  \href{http://vlado.fmf.uni-lj.si/pub/networks/data/mix/mixed.htm}{pajek} \\
 adjnoun  & UW   &  112 &  425 &        425.0 &  linguistics &  \href{http://www-personal.umich.edu/~mejn/netdata/}{Newman} \\
 afootball  & UW   &  115 &  613 &        616.0 &  social &  \href{http://www-personal.umich.edu/~mejn/netdata/}{Newman} \\
 baywet  &    &  128 &  2075 &       3459.4 &  biology &  \href{http://vlado.fmf.uni-lj.si/pub/networks/data/bio/foodweb/foodweb.htm}{pajek} \\
 jazz  & UW   &  198 &  2742 &       5484.0 &  social &  \href{http://deim.urv.cat/~aarenas/data/welcome.htm}{Arenas} \\
 SmallW\_main  & UW   &  233 &  994 &       1988.0 &  citation &  \href{http://vlado.fmf.uni-lj.si/pub/networks/data/cite/default.htm}{pajek} \\
 A01\_main  &    &  249 &  635 &        642.0 &  citation &  \href{http://vlado.fmf.uni-lj.si/pub/networks/data/}{pajek} \\
 celegansneural  &    &  297 &  2148 &       8817.0 &  biology &  \href{http://www-personal.umich.edu/~mejn/netdata/}{Newman} \\
 USAir97  & UW   &  332 &  2126 &       2126.0 &  flight &  \href{http://vlado.fmf.uni-lj.si/pub/networks/data/}{pajek} \\
 netscience\_main  &    &  379 &  914 &        489.5 &  co-author &  \href{http://www-personal.umich.edu/~mejn/netdata/}{Newman} \\
 WorldCities\_main  &    &  413 &  7518 &      16892.0 &  social &  \href{http://vlado.fmf.uni-lj.si/pub/networks/data/mix/mixed.htm}{pajek} \\
 A03  &    &  423 &  578 &        578.0 &  biology &  \href{http://vlado.fmf.uni-lj.si/pub/networks/data/GD/GD.htm}{GraphDrawing} \\
 celeg\_metab  &    &  453 &  2040 &       4596.0 &  biology &  \href{http://deim.urv.cat/~aarenas/data/welcome.htm}{Arenas} \\
 USAir500  &    &  500 &  2980 &  453914166.0 &  flight &  \href{http://cxnets.googlepages.com/}{Cx-Nets} \\
 s838  & UW   &  512 &  819 &        819.0 &  technology &  \href{http://www.weizmann.ac.il/mcb/UriAlon/}{UriAlon} \\
 Roget\_main  &    &  994 &  3641 &       5059.0 &  linguistics &  \href{http://vlado.fmf.uni-lj.si/pub/networks/data/dic/roget/Roget.htm}{pajek} \\
 SmaGri\_main  &    &  1024 &  4917 &       4922.0 &  citation &  \href{http://vlado.fmf.uni-lj.si/pub/networks/data/cite/default.htm}{pajek} \\
 A96  & UW   &  1096 &  1677 &       1691.0 &  software &  \href{http://vlado.fmf.uni-lj.si/pub/networks/data/GD/GD.htm}{GraphDrawing} \\
 email  & UW   &  1133 &  5451 &      10902.0 &  social &  \href{http://deim.urv.cat/~aarenas/data/welcome.htm}{Arenas} \\
 polBlogs\_main  &    &  1222 &  16717 &      19089.0 &  citation &  \href{http://vlado.fmf.uni-lj.si/pub/networks/data/mix/mixed.htm}{pajek} \\
 NDyeast\_main  &    &  1458 &  1993 &       1993.0 &  biology &  \href{http://vlado.fmf.uni-lj.si/pub/networks/data/ND/NDnets.htm}{pajek} \\
 Java  & UW   &  1538 &  7817 &       8032.0 &  software &  \href{http://vlado.fmf.uni-lj.si/pub/networks/data/GD/GD.htm}{GraphDrawing} \\
 Yeast\_main  &    &  2224 &  7049 &       7049.0 &  biology &  \href{http://vlado.fmf.uni-lj.si/pub/networks/data/bio/Yeast/Yeast.htm}{pajek} \\
 SciMet\_main  &    &  2678 &  10369 &      10385.0 &  citation &  \href{http://vlado.fmf.uni-lj.si/pub/networks/data/cite/default.htm}{pajek} \\
 ODLIS\_main  &    &  2898 &  16381 &      18417.0 &  linguistics &  \href{http://vlado.fmf.uni-lj.si/pub/networks/data/dic/odlis/Odlis.htm}{pajek} \\
 DutchElite\_main  & UW   &  3621 &  4310 &       4311.0 &  economy &  \href{http://vlado.fmf.uni-lj.si/pub/networks/data/2mode/DutchElite.htm}{pajek} \\
 geom\_main  &    &  3621 &  9461 &      19770.0 &  co-author &  \href{http://vlado.fmf.uni-lj.si/pub/networks/data/collab/geom.htm}{pajek} \\
 Kohonen\_main  &    &  3704 &  12675 &      12685.0 &  citation &  \href{http://vlado.fmf.uni-lj.si/pub/networks/data/cite/default.htm}{pajek} \\
 Epa\_main  & UW   &  4253 &  8897 &       8953.0 &  web &  \href{http://vlado.fmf.uni-lj.si/pub/networks/data/mix/mixed.htm}{pajek} \\
 eva\_main  &    &  4475 &  4654 &       4664.0 &  economy &  \href{http://vlado.fmf.uni-lj.si/pub/networks/data/econ/Eva/Eva.htm}{pajek} \\
 PPI\_SCerevisiae\_main  & UW   &  4626 &  14801 &      29602.0 &  biology &  \href{http://cxnets.googlepages.com/}{Cx-Nets} \\
 USpowerGrid  & UW   &  4941 &  6594 &      13188.0 &  technology &  \href{http://vlado.fmf.uni-lj.si/pub/networks/data/mix/mixed.htm}{pajek} \\
 hep-th\_main  &    &  5835 &  13815 &      13674.6 &  citation &  \href{http://www-personal.umich.edu/~mejn/netdata/}{Newman} \\
 California\_main  & UW   &  5925 &  15770 &      15946.0 &  web &  \href{http://vlado.fmf.uni-lj.si/pub/networks/data/mix/mixed.htm}{pajek} \\
 Zewail\_main  &    &  6640 &  54174 &      54244.0 &  citation &  \href{http://vlado.fmf.uni-lj.si/pub/networks/data/cite/default.htm}{pajek} \\
 Erdos02  & UW   &  6927 &  11850 &      11850.0 &  co-author &  \href{http://vlado.fmf.uni-lj.si/pub/networks/data/}{pajek} \\
 Lederberg\_main  &    &  8212 &  41436 &      41507.0 &  citation &  \href{http://vlado.fmf.uni-lj.si/pub/networks/data/cite/default.htm}{pajek} \\
 PairsP  &    &  10617 &  63786 &     612563.0 &  similarity &  \href{http://vlado.fmf.uni-lj.si/pub/networks/data/dic/fa/FreeAssoc.htm}{pajek} \\
 PGP\_main  & UW   &  10680 &  24316 &      24340.0 &  social &  \href{http://deim.urv.cat/~aarenas/data/welcome.htm}{Arenas} \\
 DaysAll  &    &  13308 &  148035 &     338706.0 &  similarity &  \href{http://vlado.fmf.uni-lj.si/pub/networks/data/CRA/terror.htm}{pajek} \\
 foldoc  &    &  13356 &  91471 &     125207.0 &  linguistics &  \href{http://vlado.fmf.uni-lj.si/pub/networks/data/dic/foldoc/foldoc.htm}{pajek} \\
 astro-ph\_main  &    &  14845 &  119652 &      33372.3 &  co-author &  \href{http://www-personal.umich.edu/~mejn/netdata/}{Newman} \\
 as-22july06  & UW   &  22963 &  48436 &      48436.0 &  web &  \href{http://www-personal.umich.edu/~mejn/netdata/}{Newman} \\
 eatRS  &    &  23219 &  305501 &     788876.0 &  linguistics &  \href{http://vlado.fmf.uni-lj.si/pub/networks/data/dic/eat/Eat.htm}{pajek} \\
 DIC28\_main  & UW   &  24831 &  71014 &      71014.0 &  linguistics &  \href{http://vlado.fmf.uni-lj.si/pub/networks/data/}{pajek} \\
 hep-th-new\_main  &    &  27400 &  352059 &     352542.0 &  co-author &  \href{http://vlado.fmf.uni-lj.si/pub/networks/data/hep-th/hep-th.htm}{pajek} \\
 cmat03\_main  &    &  27519 &  116181 &      60793.1 &  co-author &  \href{http://www-personal.umich.edu/~mejn/netdata/}{Newman} \\
 wordnet3\_main  & UW   &  75606 &  120472 &     131780.0 &  linguistics &  \href{http://vlado.fmf.uni-lj.si/pub/networks/data/}{pajek} \\
  \hline
\end{tabular}
}
  \caption{Graph collection.}
  \label{tab:graphs}
\end{table*}

\begin{table*}
  \centering
  { 
\begin{tabular}{ll}
\hline
  source & web address \\
\hline
Arenas & \url{http://deim.urv.cat/~aarenas/data/welcome.htm} \\
ANoack & \url{http://www-sst.informatik.tu-cottbus.de/~an/GD/} \\
Cx-Nets & \url{http://cxnets.googlepages.com/} \\
GraphDrawing & \url{http://vlado.fmf.uni-lj.si/pub/networks/data/GD/GD.htm} \\
Newman & \url{http://www-personal.umich.edu/~mejn/netdata/} \\
pajek & \url{http://vlado.fmf.uni-lj.si/pub/networks/data/} \\
UriAlon & \url{http://www.weizmann.ac.il/mcb/UriAlon/} \\
  \hline
\end{tabular}
 }
  \caption{Graph sources.}
  \label{tab:sources}
\end{table*}

\end{document}